\documentclass[runningheads,oribibl]{llncs}
\usepackage[latin1]{inputenc}
\usepackage{amsmath,amsfonts,amssymb}
\usepackage{theorem}
\usepackage{url}
\usepackage{verbatim}

\newcommand {\FF}{\mathbb {F}}

\newcommand {\Z}{\mathbb {Z}}
\newcommand {\F}{\mathcal {F}}
\newcommand {\C}{\mathcal {C}}
\newcommand {\M}{\mathcal {M}}
\newcommand {\Lat}{\mathcal {L}}

\newcommand {\ddiv}{\operatorname {div}}
\newcommand {\Norm}{\operatorname {N}}

\spnewtheorem{heuristic}[theorem]{Heuristic}{\bfseries}{\itshape}
\spnewtheorem{algorithm}[theorem]{Algorithm}{\bfseries}{\itshape}
\spnewtheorem{heuprop}[theorem]{Heuristic Result}{\bfseries}{\itshape}
\spnewtheorem{prop}[theorem]{Proposition}{\bfseries}{\itshape}
\spnewtheorem*{heuproof}{Justification}{\itshape}{\rmfamily}

\title {An $L (1/3 + \varepsilon)$ Algorithm for the Discrete Logarithm
Problem for Low Degree Curves}

\titlerunning{An $L (1/3 + \varepsilon)$ Algorithm for Discrete Logarithm
for Low Degree Curves}

\author {Andreas Enge\inst{1} \and Pierrick Gaudry\inst{2}}

\institute{INRIA Futurs \& Laboratoire d'Informatique (CNRS/UMR 7161)\\
École polytechnique, 91128 Palaiseau Cedex, France 
\and
LORIA (CNRS/UMR 7503), Campus Scientifique, BP 239\\
54506 Vand{\oe}uvre-lès-Nancy Cedex, France
}

\begin{document}
\maketitle

\begin{abstract}
The discrete logarithm problem in Jacobians of curves of high genus
$g$ over finite fields $\FF_q$
is known to be computable with subexponential complexity $L_{q^g}(1/2,
O(1))$. We
present an algorithm for a family of plane curves whose degrees in $X$ and $Y$
are low with respect to the curve genus, and suitably unbalanced. The finite
base fields are arbitrary, but their sizes should not grow too fast compared to
the genus.
For this family, the group structure can be computed in subexponential time of
$L_{q^g}(1/3, O(1))$, and a discrete logarithm computation takes
subexponential time of $L_{q^g}(1/3+\varepsilon, o(1))$ for any
positive~$\varepsilon$. These runtime bounds
rely on heuristics similar to the ones used in the number field sieve or the
function field sieve algorithms.
\end{abstract}

\section {Introduction}

The discrete logarithm problem in algebraic curves over finite fields has
been receiving particular attention since elliptic curves and
subsequently Jacobian groups of further algebraic curves have been
proposed for discrete logarithm based public key cryptosystems. Although
it is now clear that high genus curves are unsuitable for
cryptographical use, it remains crucial to study algorithms for solving
the discrete logarithm problem in those curves for several reasons.
The first reason is that having a better understanding of the situation
for high genus curves might lead to algorithmic improvements also in the
small genus case. The second reason is that the Weil descent
strategy of attacking the discrete logarithm problem in
elliptic curves defined over extension fields leads
to a discrete logarithm problem in the Jacobian of a high genus curve.
Therefore a better algorithm for high genus discrete logarithms becomes
naturally a potential threat for some elliptic curves.

It turned out very early that the
discrete logarithm problem in high genus hyperelliptic curves (for instance
in the sense that the size $q$ of the base field is fixed, while the genus
$g$ tends to infinity) can be solved by a subexponential algorithm of
complexity $L_{q^g}(1/2, O(1))$. The first such algorithm was proposed in \cite
{AdDeHu94}. As other subexponential algorithms, it consists of fixing a factor
base of small prime elements (here, prime divisors) and of creating
relations that correspond to the zero element modulo an equivalence
relation (here, equivalence of divisors modulo principal divisors). After
collecting sufficiently many relations and somehow introducing the base of
the discrete logarithm and the element whose logarithm is sought, linear
algebra yields the desired result.
Assuming that smooth elements, that are elements decomposing over the
factor base, have the same density as for instance smooth integers or
polynomials, such algorithms usually end up with a complexity of
$L_{q^g}(1/2, O(1))$.

The algorithm in \cite {AdDeHu94} creates relations by randomly taking low
degree functions (that are linear in $Y$ for the curve $Y^2 = f (X)$),
whose divisors are relations. Its analysis is only heuristic. The first
proven algorithms are given in \cite {MuStTh99} for the infrastructure of
real-quadratic hyperelliptic function fields and in \cite {Enge02} for
Jacobians of hyperelliptic curves. Relations are obtained in a process
similar to that of \cite {HaMc89} by taking random linear combinations of
factor base elements, reducing modulo the equivalence relation and checking
for smoothness. A rigorous analysis is derived from the lower bound on the
density of smooth divisors in \cite {EnSt02}. A generic description of a
similar algorithm can be found in \cite {EnGa02}; it applies to all class
groups in which a smoothness result is known. Heuristically, it obtains a
running time of $L_{q^g}(1/2, O(1))$ for the discrete logarithm problem in arbitrary
high genus curves, the smoothness result needed for a proof of the
complexity is however only available for hyperelliptic curves.

A proven algorithm of complexity $L_{q^g}(1/2+\varepsilon, O(1))$ for very general curves over a
fixed field $\FF_q$ and with genus $g$ tending to infinity (with the only
restriction that the curves contain a rational point and that the
cardinality of the Jacobian group is bounded by $q^{g + O (\sqrt g)}$) is
given in \cite {Couveignes01}. Unlike previous algorithms, it appears to be
specific to algebraic curves and relies on a double randomisation, taking
random combinations of factor base elements and a random function in a
Riemann--Roch space. A relation is obtained whenever the divisor of this
function is smooth.
A more general algorithm is proposed in \cite{Hess04}
that yields a proven $L_{q^g}(1/2, O(1))$ complexity without any restriction on the
input curve.

Another line of research on the discrete logarithm problem for algebraic
curves, started in \cite {Gaudry00} and not pursued in this article,
consists of fixing $g$ and having $q$ tend to infinity. This leads to
algorithms that are exponential, but faster than generic algorithms of
square root complexity as soon as $g \geq 3$, see \cite {GaThThDi07,Diem06}.

In the light of algorithms of complexity $L(1/3)$ for the discrete
logarithm problem in finite fields as well as for factoring integers, it has
been an open problem to determine whether this complexity can be achieved
also for algebraic curves. In this article, we present the first
probabilistic algorithm of heuristic complexity $L_{q^g}(1/3, O(1))$ to
compute the group structure of certain
curves whose total degree is relatively small compared to their genus. When
introducing the two elements of the Jacobian for which the discrete
logarithm problem is to be solved, some sacrifice has to be made; we obtain
an algorithm of complexity bounded by
$L_{q^g}(1/3 + \varepsilon, o(1))$ for any positive constant
$\varepsilon$.

The relation collection phase is the same as in \cite {AdDeHu94} and
consists of looking for smooth divisors of functions linear in $Y$.
By applying it to the curves of our special family, one readily obtains a lower
degree of the affine part of the intersection divisor than in the general case,
from which a complexity of $L_{q^g}(1/3, O(1))$ is derived.
For smoothing the two divisors involved in the discrete
logarithm problem, a process is employed that is similar to the one used
in the number field sieve or in the function field sieve. This is the
general {\em special-$Q$ descent} strategy (also related to the so-called
lattice sieving).  Each divisor is partially smoothed into prime divisors of
degree less than the starting divisor. Then each such prime divisor $Q$ is
smoothed again into smaller prime divisors, and we iterate until every
divisor is rewritten in terms of elements of the factor base. 
However, in our case it is necessary to add an arbitrarily small
constant $\varepsilon$ to the $1/3$ parameter to obtain a proper descent
phenomenon; otherwise, the process would get stuck after one step.

Let us mention that subsequently to our algorithm, Diem has presented at
the 10th Workshop on Elliptic Curve Cryptography (ECC 2006) an algorithm based
on similar ideas, but with a quite different point of view. He manages to
obtain a complexity of $L (1/3, O (1)$ for the discrete logarithm phase, for
which our algorithm takes $L (1/3 + \varepsilon, o (1))$. We will show how to
reach a complexity of $L (1/3, O (1))$ for discrete logarithms in our setting
in the long, journal version.

\paragraph{Acknowledgement.} We thank Claus Diem for his careful reading of
our article and many useful remarks. 

\section {Main idea}

Before describing our algorithm with all its technical details on a
general class of curves, we sketch in this section the main idea yielding a
complexity of $L_{q^g}(1/3, O(1))$ for the relation collection
phase for a restricted class of curves. We provide a simplified analysis by
hand waving; Section~\ref {sec:smoothness} is devoted to a more precise
description of the heuristics used and of the smoothness properties needed
for the analysis.

Let $\FF_q$ be a fixed finite field. We consider a family of $C_{ab}$ curves
over $\FF_q$, that is, curves of the form
\[
\C : Y^n + X^d + f (X, Y)
\]
without affine singularities such that
$\gcd (n, d) = 1$ and any monomial $X^i Y^j$ occurring in $f$ satisfies
$n i + d j < nd$.
Such a curve has genus $g = \frac {(n-1)(d-1)}{2}$; we assume that $g$
tends to infinity, and that $n \approx g^{1/3}$ and $d \approx g^{2/3}$
(we use the symbol $\approx$, meaning ``about the same size''
with no precise definition).
The non-singular model of a $C_{ab}$ curve has a unique point at infinity,
and it is $\FF_q$-rational; so there is a natural bijection between degree
zero divisors and affine divisors, and in the following, we shall only be
concerned with effective affine divisors. Choose as factor base $\F$ the
$L_{q^g}(1/3, O(1))$ prime divisors of smallest degree (that is, the prime
divisors up to a degree of $B \approx \log_q L_{q^g}(1/3, O(1))$).
To obtain relations, consider functions linear in $Y$ of the form
\[
\varphi = a (X) + b (X) Y
\]
with $a$, $b \in \FF_q [X]$, $\gcd (a, b) = 1$ and $\deg a$, $\deg b =
\delta \approx g^{1/3}$.
Whenever the affine part $\ddiv (\varphi)$ of the divisor of $\varphi$ is
smooth with respect to the factor base, it yields a relation, and we have
to estimate the probability of this event.

Let $\Norm$ be the norm of the function field extension
$\FF_q (\C) = \FF_q (X)[Y] / (Y^n + X^d + f (X, Y))$ relative to $\FF_q (X)$.
The norm of $\varphi$ is computed as
\begin {eqnarray*}
\Norm (\varphi) & = & \Norm (b) \Norm \left( Y + \frac {a}{b} \right) \\
& = & b^n \left( \left( -\frac {a}{b} \right)^n + X^d
   + f \left( X, - \frac {a}{b} \right) \right) \\
& = & (-a)^n + b^n X^d + f^* (X),
\end {eqnarray*}
where each monomial $X^i Y^j$ occurring in $f$ is transformed into a
monomial $X^i (-a)^j b^{n-j}$ in~$f^*$.

Since $\varphi$ is linear in $Y$, all prime divisors it contains are
totally split over $\FF_q (X)$, and $\varphi$ is $B$-smooth if and only if
its norm is.
We have
\[
\deg_X \Norm (\varphi)
\leq \max (n \deg a, n \deg b + d)
= n \delta + d
\approx g^{2/3}.
\]
Heuristically, we assume that the norm behaves like a random polynomial of
degree about $g^{2/3}$. Then it is $B$-smooth with probability
$1 / L_{q^g}(1/3, O(1))$ (this is the same theorem as the one stating that a random
polynomial of degree $g$ is $\log_q L_{q^g}(1/2, O(1))$-smooth with probability
$1 / L_{q^g}(1/2, O(1))$, cf., for instance, Theorem~2.1 of \cite {BePo98}).
Equivalently, we may observe that $\deg (\ddiv (\varphi)) = \deg_X (\Norm
(\varphi))$ and assume heuristically that $\ddiv (\varphi)$ behaves like a
random effective divisor of the same degree. Then the standard results on
arithmetic semigroups (cf. Section~\ref {sec:smoothness}) yield again that
$\ddiv (\varphi)$ is smooth with probability $1 / L_{q^g}(1/3, O(1))$.

Thus, the expected time for obtaining $|\F| = L_{q^g}(1/3, O(1))$ relations is
$L_{q^g}(1/3, O(1))$, which is also the complexity of the linear algebra step for
computing the Smith normal form and thus the group structure of the
Jacobian. The complexity of the discrete logarithm problem is not
considered here, an analysis for the full algorithm is given in
Section~\ref {sec:logarithms}.

It remains to show that the search space is sufficiently large to yield the
required $L_{q^g}(1/3, O(1))$ relations, or otherwise said, that the number of
candidates for $\varphi$ is at least $L_{q^g}(1/3, O(1))$. The number of
$\varphi$ is about
\begin {eqnarray*}
q^{2 \delta} & = & q^{2 g^{1/3}} = \exp (2 \log q g^{1/3}) \\
& < & \exp (2 (g^{1/3} (\log q)^{1/3}) (\log (g \log q))^{2/3})
= L_{q^g}(1/3, O(1)).
\end {eqnarray*}

The previous inequality in the place of the desired equality shows that a
more rigorous analysis requires a more careful handling of the $\log q$
factors; in particular, $\delta$ has to be slightly increased. Moreover,
the constant exponent in the subexponential function needs to be taken into
account. This motivates the following section, in which we examine in more
detail the smoothness heuristics and results that are needed for the
algorithm.

\section {Smoothness}
\label {sec:smoothness}

The algorithm presented in this article relies on finding relations as
smooth divisors of random polynomial functions of low degree. We suppose
that all curves are given by an absolutely irreducible plane affine model
\[
\C : F (X, Y)
\]
with $F \in \FF_q [X, Y]$, where $\FF_q$ is the exact constant field of
the function field of $\C$.
The factor base $\F$ consists essentially of the places of degree bounded by
some parameter $\mu$, with a few technical modifications. Precisely, $\F$ is
composed of the following places:
\begin {itemize}
\item
the places corresponding to the resolution of singularities, regardless of
their degrees, whose number is bounded by $\frac {(d-1)(d-2)}{2}$ with $d =
\deg F$. By including them in $\F$, the algorithm can be described as if the curves
were non-singular.
\item
the infinite places corresponding to non-singularities, regardless of their
degrees, whose number is bounded by $d$ by Bézout's theorem. By adding them,
it becomes sufficient to only examine the affine part of any divisor.
\item
places of degree bounded by some parameter $\mu$ and of inertia degree $1$
with respect to the function field extension $\FF_q (X)[Y] / (F)$ over $\FF_q
(X)$. Otherwise said, places corresponding to prime ideals of the form $(u, Y
- v)$ with $u \in \FF_q [X]$ irreducible of degree at most $\mu$ and $v \in
\FF_q [X]$ of degree less than $\deg u$; the inertia degree is in fact the
degree of the second generator in $Y$. Due to the way relations are obtained
in the algorithm, no places of higher inertia degree may occur.
\end {itemize}

A divisor is called $\F$-smooth if it can be decomposed over the factor
base; thus only its affine part plays a role, and for polynomial functions,
this is an effective (i.e. non-negative) divisor. An effective divisor is
called $\mu$-smooth if it is composed only of places of degree up to $\mu$.
To be able to analyse the smoothness probability, we need the
following reasonable assumption.

\begin {heuristic}
\label {heu1}
Let $D$ be the divisor of a uniformly randomly chosen polynomial of the form
$b (X) Y - a (X)$ and $\nu$ the degree of its affine part. Then the
probability of $D$ to be $\F$-smooth is the same as that of a random effective
divisor of degree $\nu$ to be $\mu$-smooth.
\end {heuristic}

Heuristic~\ref {heu1} covers the relation collection phase. For computing
discrete logarithms, arbitrary non-principal divisors need to be smoothed, and
another assumption is needed.

\begin {heuristic}
\label {heu2}
The probability of a uniformly randomly chosen effective divisor of degree
$\nu$ to be $\F$-smooth is essentially the same as that of being $\mu$-smooth.
\end {heuristic}

Heuristic~\ref {heu2} claims in fact that places of inertia degree larger
than~$1$ do not play a role for smoothness considerations.
In the analogous case
of number fields this is justified by the observation that these places have a
Dirichlet density of~$0$, and the situation is completely analogous for
function fields: A place of degree $\mu$ and inertia degree $f$ dividing $\mu$
corresponds to a closed point on $\C$ with $X$-coordinate in $\FF_{q^{\mu/f}}$
and $Y$-coordinate in $\FF_{q^\mu}$, of which there are on the order of
$q^{\mu/f}$. Clearly, places with $f \geq 2$ are completely negligible.

The probability of $\mu$-smoothness is ruled by the usual results on
smoothness probabilities in arithmetic semigroups such as the integers or
polynomials over a finite field, cf. \cite {Manstavicius92}.

Unfortunately, most results in the literature assume a fixed semigroup and
give asymptotics for $\mu$ and $\nu$ tending to infinity, whereas we need
information that is uniform over an infinite family of curves. Theorem~13
of \cite {Hess04} provides such a result:

\begin {theorem}[He\ss]
\label {th:hess}
Let $0 < \varepsilon < 1$, $\gamma = \frac {3}{1 - \varepsilon}$ and $\nu$,
$\mu$ and $u = \frac {\nu}{\mu}$ such that
$3 \log_q (14 g + 4) \leq \mu \leq \nu^\varepsilon$ and
$u \geq 2 \log (g + 1)$.
Denote by $\psi (\nu, \mu)$ the number of $\mu$-smooth effective divisors of
degree $\nu$. Then for $\mu$ and $\nu$ sufficiently large (with an explicit
bound depending only on $\varepsilon$, but not on $q$ or $g$),
\[
\frac {\psi (\nu, \mu)}{q^\nu} \geq e^{- u \log u \left( 1 +
\frac {\log \log u + \gamma}{\log u} \right)}
= e^{- u \log u (1 + o (1))}.
\]
\end {theorem}

Notice that the proof of Theorem~\ref {th:hess}, similar in spirit to that for
hyperelliptic curves in \cite{EnSt02}, is entirely combinatorial and relies on
the fact that there are essentially $q^\mu / \mu$ places of degree $\mu$.
So we expect the result to hold even if one restricts to places of inertia
degree~$1$.

Denote by
\[
L (\alpha, c) = L_{q^g} (\alpha, c)
= e^{c (g \log q)^\alpha (\log (g \log q))^{1 - \alpha}}
\]
for $0 \leq \alpha \leq 1$ and $c > 0$
the subexponential function with respect to $g \log q$, and
let
\[
\M = \M_{q^g} = \log_q (g \log q) = \frac {\log (g \log q)}{\log q}.
\]
The parameter $g \log q$ will be the input size for the class of curves
we consider; more intrinsically, this is
the logarithmic size
of the group in which the discrete logarithm problem is defined.

\begin {prop}
\label {prop:smoothness}
Let
$
\nu = \lfloor \log_q L (\alpha, c) \rfloor
    = \lfloor c g^\alpha \M^{1 - \alpha} \rfloor
$
and
$
\mu = \lceil \log_q L (\beta, d) \rceil
    = \lceil d g^\beta \M^{1 - \beta} \rceil
$
with $0 < \beta < \alpha \leq 1$ and $c$, $d > 0$.
Assume that there is a constant
$\delta > \frac {1 - \alpha}{\alpha - \beta}$ such that
$g \geq (\log q)^\delta$.
Then for $g$ sufficiently large,
\[
\frac {\psi (\nu, \mu)}{q^\nu} \geq L \left( \alpha - \beta, - \frac {c}{d}
(\alpha - \beta) + o (1) \right),
\]
where $o (1)$ is a function that is bounded in absolute value by a constant
(depending on $\alpha$, $\beta$, $c$, $d$ and $\delta$) times
$\frac {\log \log (g \log q)}{\log (g \log q)}$.
\end {prop}

\begin {proof}
One computes
\[
u = \frac {\nu}{\mu}
  \leq \frac {c}{d} \left( \frac {g \log q}{\log (g \log q)}
  \right)^{\alpha - \beta}
\]
(the inequality being due only to the rounding of $\nu$ and $\mu$),
\[
\log u = (\alpha - \beta) \log (g \log q) (1 + o (1))
\]
and 
\[
\frac {\log \log u}{\log u} = o (1),
\]
with both $o (1)$ terms being of the form stipulated in the proposition.
Applying Theorem~\ref {th:hess} yields the desired result. Its
prerequisites are satisfied since
\begin {eqnarray*}
\overline \lim_{g \to \infty} \frac {\log \mu}{\log \nu}
& = & \overline \lim_{g \to \infty} \frac {\beta \log g - (1 - \beta) \log
  \log q}{\alpha \log g - (1 - \alpha) \log \log q} \\
& \leq & \overline \lim_{g \to \infty} \frac {\beta \log g}{\alpha \log g -
\frac {1 - \alpha}{\delta} \log g} \\
& = & \frac {\beta}{\alpha - \frac {1 - \alpha}{\delta}}
=: \varepsilon < 1
\end {eqnarray*}
because of the definition of $\delta$.
Notice further that $g \to \infty$ is equivalent to $g \log q \to \infty$,
and that also $\mu$ and $\nu$ tend to infinity when $g$ does.
\hfill \qed
\end {proof}

The choice of $\mu$ shall insure that the factor base size, that is about
$q^\mu$, becomes subexponential. But the necessary rounding of $\mu$, which
may increase $q^\mu$ by a factor of almost $q$, may result in more than
subexponentially many elements in the factor base when $q$ grows too fast
compared to~$g$.

\begin {prop}
\label {prop:subexponentiality}
Let $0 < \beta < 1$ and $\delta > \frac {1 - \beta}{\beta}$.
If $g \geq (\log q)^\delta$, then
$q = L (\beta, o (1))$ for $g \to \infty$.
In particular,
$\delta > \max \left( \frac {1 - \alpha}{\alpha - \beta}, \frac {1 -
\beta}{\beta} \right)$ in Proposition~\ref {prop:smoothness} implies that
$q^\mu = L (\beta, d + o (1))$.
\end {prop}

\begin {proof}
To verify the first assertion, one computes
\begin {eqnarray*}
q & = & e^{\log q} = e^{(\log q)^{1 - \beta} (\log q)^\beta} \\
& \leq & e^{g^{(1 - \beta) / \delta} (\log q)^\beta (\log (g \log q))^{1 -
\beta}} \\
& = & e^{(g \log q)^\beta (\log (g \log q)^{1 - \beta})
g^{\frac {1 - \beta}{\delta} - \beta}},
\end {eqnarray*}
and $g^{\frac {1 - \beta}{\delta} - \beta} \to 0$
since $\frac {1 - \beta}{\delta} - \beta < 0$.
The second assertion is obvious.
\hfill \qed
\end {proof}

\section {Computing the group structure}
\label{sec:grp}

This section is concerned with the relation collection phase of the discrete
logarithm algorithm; an immediate application is the computation of the
cardinality and the group structure of the Jacobian of the curve.
Relation collection is virtually identical to the process described for
hyperelliptic curves in \cite {AdDeHu94}; the running time of $L (1/3,
O(1))$ is
obtained by applying it to a particular class of curves that are of relatively
low degree with respect to their genus and for which the degrees in $X$ and
$Y$ of a plane model are balanced in a certain way.

We consider absolutely irreducible curves over finite fields $\FF_q$ of
characteristic $p$ of the form
\[
\C : Y^n + F (X,Y)
\]
with $F (X, Y) \in \FF_q [X]$ of degree $d$ in $X$ and at most $n-1$ in $Y$.
The function field extension $\FF_q (\C) = \FF_q (X)[Y] / (Y^n+F(X,Y))$ over $\FF_q
(X)$ is supposed to be separable (which is for instance the case if
$p \nmid n$).

Most importantly, the degrees $n$ and $d$ are related to the genus $g$ by
\[
n \leq n_0 g^{1/3} \M^{-1/3} \text { and } d \leq d_0 g^{2/3} \M^{1/3}
\]
where $\M = \frac {\log (g \log q)}{\log q}$ and
$n_0$, $d_0$ are some positive constants.

For instance, $\C$ may be a $C_{ab}$ curve of degree $n \sim g^{1/3}
\M^{-1/3}$ in $Y$ and $d \sim 2 g^{2/3} \M^{1/3}$ in $X$.

For the running time analysis, we will want to apply Propositions~\ref
{prop:smoothness} and \ref {prop:subexponentiality} with $\alpha = 2/3$ and
$\beta = 1/3$; so we have to assume that the curves belong to a family
satisfying $g \geq (\log q)^\delta$ for some $\delta > 2$.

\goodbreak
\begin {algorithm}[Group structure]\label{algo:grp} \\
\textbf {Input:} a curve $\C$ as above

\noindent
\textbf {Output:} $h = |J_{\C} (\FF_q)|$ and divisors
   $D_1, \ldots, D_r$ with their orders $h_1, \ldots, h_r$
   s.t. $J_{\C} (\FF_q) = \langle D_1 \rangle \times \cdots \times
   \langle D_r \rangle$

\begin {enumerate}
\item
Compute an approximation of $h$ within a factor of $2$, that is,
$h_-$ and $h_+$ s.t.
\[
h_- < h < h_+ \text { and } h_+ \leq 2 h_-.
\]
\item
Fix a smoothness bound $B = \lceil \log_q L (1/3, \rho) \rceil$
(with a parameter $\rho$ to be determined later) and compute the factor base
$\F$ consisting of all affine prime divisors of $\C$ of degree at most $B$ as
well as all infinite prime divisors
and prime divisors corresponding to singularities
regardless of their degrees. Let $t =
|\F|$ and $\F = \{ P_1, \ldots, P_t \}$.
\item
Start with an empty matrix of relations $R$ and repeat the following step
until $s \ge 2t$ relations are obtained (in practice, $s$ slightly larger
than $t$ should suffice):

Draw uniformly at random a function
\[
\varphi = b (X) Y - a(X) \in \FF_q (\C)
\]
with $a$, $b \in \FF_q [X]$ of degree at most
\[
m = \lfloor \sigma g^{1/3} \M^{2/3} \rfloor
\]
(with a parameter $\sigma$ to be determined later). If its divisor is
$\F$-smooth, that is,
\[
\ddiv \varphi = \sum_{i=1}^t e_i P_i,
\]
add a column $(e_1, \ldots, e_t)^T$ to the matrix $R$.
\item
Compute the rank of $R$; if it is less than $t$, declare failure and stop.
\item
Compute the Smith normal form $S = \operatorname {diag} (h_r, \ldots, h_1, 1,
\ldots, 1)$ of $R$, where $1 \neq h_1 | h_2 | \cdots | h_r$, and
unimodular transformation matrices $T \in \Z^{t \times t}$ and
$U \in \Z^{s \times s}$ s.t.
$T R U = (S | 0)$.

Let $h = h_1 \cdots h_r$. If $h \geq h_+$, declare failure and stop.

Otherwise return $h$, $D_1, \ldots, D_r$ s.t.
\[
(D_1, \ldots, D_r, 0, \ldots, 0) = (P_1, \ldots, P_t) \, T^{-1}
\]
and $h_1, \ldots, h_r$.
\end {enumerate}
\end {algorithm}

That the algorithm is correct follows from standard arguments such as given in
\cite{AdDeHu94,Enge02,EnGa02}. It remains to
prove its failure probability and running time. We also have to show that
there actually are subalgorithms to carry out the different steps; these are
given together with the following running time analysis.

\begin {enumerate}
\item
An approximation $\tilde h$ of $h$ can be obtained by appropriately truncating
the $L$-series of the curve as in \cite [Section~6]{Hess04}. The necessary
counting of the number of points on the curve over a small number of extension
fields is shown in \cite {Hess04} to be polynomial in $g$ and $\log q$
for curves of degree in $O (g)$. The bounds on $h$ are then given by $h_- =
\tilde h / \sqrt 2$ and $h_+ = \sqrt 2 \tilde h$.
\item
The affine prime divisors of degree up to $B$ are obtained by enumerating all
irreducible monic polynomials $f \in \FF_q [X]$ of degree up to $B$ and
factoring
$Y^n + F (X, Y)$ over
$\FF_q [X] / (f) [Y]$. Each factor of degree $w$ yields a prime divisor of
degree $w \deg f$.
Altogether, these factorisations can be carried out by $O (q^B)$ repetitions
of a randomised algorithm with an expected running time that is polynomial in
$n$, $B$ and $\log q$, and thus ultimately in $g \log q$. Since polynomial
terms are in $L (1/3, o (1))$, they can be neglected, and we retain only the
term $O (q^B)$ for the remainder of the analysis.

The number of singular places is bounded by $O ((n d)^2) = O(g^2)$ using the
genus formula for a plane curve. They can be fully described in polynomial
time, by computing the desingularisation trees of the singular points
(see for instance~\cite{Hache96}).

The non-singular places at infinity are included in the intersection of the
projective curve with the line $Z = 0$, which has at most $O (nd) = O (g)$
elements by Bézout's theorem, and these are also computable in polynomial
time.

So this step terminates with a factor base of size
\[
t = O \left( n q^B \right) = L (1/3, \rho + o (1))
\]
that is computed in time $L (1/3, \rho + o (1))$.
\item
To estimate the smoothness probability of $\ddiv \varphi$ under Heuristic~\ref
{heu1}, we need to compute the degree of its affine part. Denote the affine
degree of a divisor by $\deg_\text {aff}$.
Let $\sigma_1, \ldots, \sigma_n$ be the different embeddings of $\FF_q (\C)$
into its Galois closure (that exists because the function field extension is
assumed to be separable). The $\sigma_i$ fixing $\FF_q (X)$, they send affine
to affine and infinite to infinite prime divisors. Hence, all the
$\deg_\text {aff} (\varphi^{\sigma_i})$ are the same and given by
\[
\deg_\text {aff} \varphi
= \frac {1}{n} \deg_\text {aff} \Norm_{\FF_q (\C) / \FF_q (X)} (\varphi)
= \deg_X \Norm (\varphi).
\]
The norm of $\varphi$ is computed as
$\Norm (\varphi) = \operatorname {Res}_Y (\varphi, Y^n+F(X,Y))$, and its degree in $X$
is bounded from above by
\[
\deg_X \varphi \cdot \deg_Y \C + \deg_Y \varphi \cdot \deg_X \C
= n m + d.
\]
The divisor of $\varphi$ is $B$-smooth if and only if its norm is; this test
as well as the decomposition of a smooth $\ddiv \varphi$ into prime divisors
boils down to a factorisation of the norm in $\FF_q [X]$ and takes random
polynomial time.

Let $\tau = (n_0 \sigma + d_0) / 3$.
Applying Propositions~\ref {prop:smoothness} and \ref {prop:subexponentiality}
under Heuristic~\ref {heu1} with
$n m + d \leq 3 \tau g^{2/3} \M^{1/3}$ in the place of $\nu$
and $B = \lceil \rho g^{1/3} \M^{2/3} \rceil$ in the place of $\mu$ shows that
a relation is obtained on average in time
$
L \left( 1/3, \frac {\tau}{\rho} + o (1) \right),
$
so that this step takes overall
\[
L \left( 1/3, \frac {\tau}{\rho} + \rho + o (1) \right).
\]
\item and 5.
Since all entries of the matrix are of bit size polynomial in $g \log q$, its
rank and Smith normal form can be computed in quartic time according to
\cite[Proposition~8.10]{Storjohann00}, that is in
\[
L (1/3, 4 \rho + o (1)).
\]
\end {enumerate}

The total running time of the algorithm thus becomes
\[
L \left( 1/3, \max \left( \frac {\tau}{\rho} + \rho, 4 \rho \right) + o (1)
\right)
\]
with
$\tau = (n_0 \sigma + d_0) / 3$.

For any fixed $\sigma$ (and thus $\tau$), the value of $\rho$ that
minimises the running time is  $\rho = \sqrt {\tau / 3}$ and we get a
complexity of $L \left( 1/3, \frac {4 \sqrt \tau}{\sqrt 3} + o (1) \right)$.

Now $\tau$ is not a completely free parameter; it is connected to the success
probability of the algorithm.
It is in fact not clear whether the algorithm has a non-zero success
probability at all; as in \cite {AdDeHu94}, it is already unknown whether the
principal divisors of the special form considered in Step~3. generate the full
relation lattice. The analysis of the proven subexponential algorithm in \cite
{Enge02}, for instance, exploits the fact that the created relations are
essentially uniformly distributed among all possible relations in a hypercube
of side length about $|J_{\C} (\FF_q) |$. Since all our relations are sparse,
this line of argumentation definitely cannot be applied; as in \cite
{AdDeHu94},
the non-negligible success probability of the algorithm can only be
conjectured (and notice also that it does not follow from a smoothness
assumption such as Heuristic~\ref {heu1}).

A necessary condition for the success of the algorithm is nonetheless that the
number of potential functions $\varphi$ tested for smoothness in Step~3. must
be at least as large as the number of tests, since otherwise the matrix is
filled with redundant multiple relations. Thus we need
$q^{2m} \geq L \left( 1/3, \frac {4 \sqrt \tau}{\sqrt 3} \right)$ or, taking
logarithms,
\[
2 \sigma \geq \frac {4}{\sqrt 3} \sqrt \tau = \frac {4}{3}
\sqrt {n_0 \sigma + d_0},
\]
which holds asymptotically for $\sigma \to \infty$. Precisely, the optimal
value of $\sigma$ is the positive solution of the quadratic equation
$\sigma^2 - \frac {4}{9} n_0 \sigma - \frac {4}{9} d_0 = 0$.

\section{Computing discrete logarithms}
\label{sec:logarithms}

In order to smooth the basis of the discrete logarithm and the element whose
logarithm is sought, we are going to perform a special-Q descent with a
slightly larger subexponentiality parameter $1/3+\varepsilon$. Let us first
describe an algorithm that does one step of the special-Q descent and that
will be used as a building block by the final algorithm.

\begin{heuprop}\label{prop:descent}
Let $Q$ be an affine prime divisor of
the curve
$\C$ of the form
$\ddiv(u(X),Y-v(X))$, with $\deg u(X)\le \log_q L(1/3+t, c)$ for some
constants $c>0$ and $\varepsilon < t\le 1/3-\varepsilon$. 
There is an algorithm that finds a divisor $R$ equivalent to $Q$ such that all
prime divisors of $R$ are either in $\F$ or have a degree bounded by $\log_q
L(1/3+t-\varepsilon, c')$, and such that all these
prime divisors are of the form $\ddiv (u_i(X), Y - v_i(X))$.
The heuristic expected running time is bounded by
$L(1/3+\varepsilon, \frac{cn_0}{c'}(1/3+\varepsilon + o(1)))$.
\end{heuprop}

\begin{heuproof}
Let us consider the set $\Lat_Q$ of functions of the form $a(X) + b(X)Y$ whose
divisors contain $Q$ in their support. In other words, this is the
$\FF_q[X]$-lattice
$$ \Lat_Q = \{ a(X) + b(X)Y\ :\ u(X) | a(X) + v(X)b(X) \}.$$
A basis of this lattice is given by the two vectors $b_1 = u(X)$
and $b_2 = -v(X) + Y$.
 Hence,
$$ \Lat_Q = \{ \lambda(X) b_1 + \mu(X) b_2\ : \ \lambda, \mu \in \FF_q[X]
\}.$$
When $\lambda$ and $\mu$ are taken of degree at most $\delta = \log_q
L(1/3+t, c)$, the
function $\varphi$ corresponding to $\lambda(X) b_1 + \mu(X) b_2$ has the
form $a(X)+b(X)Y$ with $a$ and $b$ of degree $\Delta \le 2\log_q
L(1/3+t, c)$. The degree of the norm of $\varphi$ is then $\Delta n + d$,
which is dominated by $\log_q L(2/3+t, cn_0)$.

We rely now on Heuristic~\ref{heu1} that says that the zero divisor of the
function has the same smoothness properties as a random effective
divisor of the same degree, and apply Proposition~\ref{prop:smoothness}. 
Therefore the expected number of functions one has to try before having
found one whose divisor is $\log_q L(1/3+t-\varepsilon, c')$-smooth is 
$$ L\left(1/3+\varepsilon, \frac{cn_0}{c'}(1/3+\varepsilon + o(1))\right).$$

The fact that the prime divisors that we obtain are of the same
form as $Q$ comes from the shape of the function we have chosen.

It remains to check that the number of functions we can test in the
lattice is large enough compared to this expected number of tests. With
our choice of $\delta$, the size of the sieving space is $L(1/3+t, 2c)$, which
is larger than any $L(1/3+\epsilon)$ since $t$ is greater than
$\varepsilon$.
\hfill \qed
\end{heuproof}

This result suffices to carry out a full descent if one can initialise
the process and finish it once smoothness is reached up to a
$t<\varepsilon$. The next two heuristic results explain these steps.

\begin{heuprop}\label{prop:finaldescent}
Assume that $\rho>(\frac13+\varepsilon)\frac{n_0}{2}$.
Let $Q$ be an affine prime divisor of $\C$ of the form $\ddiv (u (X),
Y-v(X))$, with $\deg u(X)\le \log_q L(1/3+t, c)$, for some
constants $c>0$ and $0 < t\le \varepsilon$.
There is an algorithm that finds a divisor $R$ equivalent to $Q$ such
that all prime
divisors of $R$ are in $\F$ (defined with this value of $\rho$),
and such that all these
prime divisors are of the form $\ddiv (u_i(X), Y - v_i(X))$. The
heuristic expected running time is bounded by
$L\left(1/3+t, (1/3+t)\frac{cn_0}{\rho}+o(1)\right).$
\end{heuprop}

\begin{heuproof}
Let us consider the same lattice $\Lat_Q$ as in the proof of
Proposition~\ref{prop:descent}. Assume that $\lambda$ and $\mu$ are taken
of degree at most $\delta = \log_q L(1/3 + t, c)$, then, as before, the
norm of the corresponding functions are of degree bounded by $\log_q
L(2/3 + t, cn_0)$. Using again Heuristic~\ref{heu1}, one gets by
Proposition~\ref{prop:smoothness} that a $\log_q L(1/3, \rho)$-smooth
divisor can be obtained in heuristic expected time
$$ L\left(1/3 + t, (1/3+t)\frac{cn_0}{\rho}+o(1)\right).$$

One has to check that we have enough possibilities for $\lambda$ and
$\mu$ to cover this search. The sieving space is $q^{2\delta} = L(1/3+t,
2c)$. Therefore it is large enough if $2c > (1/3+t)\frac{cn_0}{\rho}$,
that is if $\rho > (1/3+t)\frac{n_0}{2}$. Since $\varepsilon>t$, this is
guaranteed by our hypothesis on $\rho$.
\hfill \qed
\end{heuproof}

\begin{heuprop}\label{prop:hm}
Let $D$ be a degree 0 divisor and $\sum_P e_P P$ its decomposition into prime
divisors such that $\sum_P |m_P| \in O (g)$.
Then there is an algorithm that finds a divisor $R$ equivalent to $D$ such
that all prime divisors of $R$ are of the form $\ddiv (u_i(X), Y -
v_i(X))$ with $\deg u_i(X) \le \log_q L(2/3-\varepsilon, c)$. 
The heuristic expected running time is bounded by
$L(1/3+\varepsilon, (1/3+\varepsilon)\frac1c+o(1))$.
\end{heuprop}

\begin{heuproof}
In order to smooth $D$, we apply the classical Hafner-McCurley strategy: a
random linear combination of elements of the factor base is added to $D$,
and the obtained divisor is tested for smoothness. Each test takes
polynomial time since the effective group law in the Jacobian reduces to
computing Riemann-Roch spaces as in \cite{Hess02}. 

Following Heuristic~\ref{heu2}, the additional restriction on the form of the
prime divisors has no influence on the running time, and the desired result
follows from Proposition~\ref{prop:smoothness}.
\hfill \qed
\end{heuproof}

Armed with these heuristic partial smoothing results, we can now derive a full
special-Q descent algorithm. Let us fix a constant $\varepsilon>0$, a
parameter of the algorithm. This $\varepsilon$ is to be thought of as small
(and of course $\varepsilon<1/6$). The algorithm assumes that
Algorithm~\ref{algo:grp} has been run as a precomputation, 
with a value
of $\rho$ that is larger than a bound given below. Similarly, the
constants $c_0$ and $c_K$ are made explicit below.

\begin{algorithm}[Discrete logarithm]
\label {alg:dlog}
\begin{enumerate}
\item Use Heuristic Result~\ref{prop:hm} to build a list $L$ of prime
divisors of degree at most $\log_q L(2/3-\varepsilon, c_0)$, such that if
we know their discrete logarithms, the discrete logarithm of $D$ is implied.
\item While there is a $Q$ in $L$ of degree more than $\log_q
L(1/3+\varepsilon,c_K)$, use Heuristic Result~\ref{prop:descent} to replace $Q$
in $L$ by a list of prime divisors of degree bounded by a subexponential
function with parameter reduced by $\varepsilon$.
\item For each $Q$ in $L$ that is not in $\F$, use
Heuristic Result~\ref{prop:finaldescent} to decompose $Q$ in~$\F$.
\end{enumerate}
\end{algorithm}

In order to analyse the algorithm, let us model it by a tree: the root is
the divisor $D$, its sons are the prime divisors coming from its
decomposition using Heuristic Result~\ref{prop:hm}, then each internal node
corresponds to a prime divisor and its sons are the prime divisors
obtained using Heuristic Result~\ref{prop:descent} or
Heuristic Result~\ref{prop:finaldescent}. The depth of the tree is
bounded by $1/ (3\varepsilon)$ since at each intermediate step the
subexponential
parameter is reduced by at least $\varepsilon$ and one has to cover a
range of $1/3$. The number of sons of each node is bounded by $g$. Hence
the total number of nodes is bounded by $g^{1/ (3\varepsilon)}$. Since
$\varepsilon$ is a fixed constant, this is a polynomial in $g\log q$ and
therefore contributes only for a $o(1)$ in the subexponential complexity.

Let us allow a computation time of $L(1/3+\varepsilon, \nu+o(1))$, for
fixed positive constants $\varepsilon$ and $\nu$. Then the first step that uses
Heuristic Result~\ref{prop:hm} can decompose $D$ in prime divisors of degree
at most $\log_q L(2/3-\varepsilon, c_0)$ in time $L(1/3+\varepsilon,
\nu + o(1))$ for $c_0 = (1/3+\varepsilon)/\nu$. Going one step down the tree,
one can decompose these primes using Heuristic Result~\ref{prop:descent} in
primes of degrees at most $\log_q L(2/3-2\varepsilon, c_1)$ in the same
time, for $c_1 = c_0n_0(1/3+\varepsilon)/\nu$. Going from level $k$ to
level $k+1$ in the tree will decompose in primes of degree at most
$\log_q L(2/3-(k+2)\varepsilon, c_{k+1})$ in the same
time, for $c_{k+1} = c_{k}n_0(1/3+\varepsilon)/\nu$. Finally, each last
step will be feasible in the same running time if
$\rho>c_{K}n_0(1/3+\varepsilon)/\nu$, where $K$ is the depth of the tree.

This value of $\rho$ is feasible and does not affect the overall
complexity. It only changes the exponent in the $L(1/3)$ runtime
of the group structure algorithm, whose complexity remains negligible
compared to the $L(1/3+\varepsilon)$ of the present algorithm.
Therefore, a suitable choice of $\rho$, $c_0$ and $c_K$ in Algorithm~\ref
{alg:dlog} results in a running time of $L (1/3 + \varepsilon, \nu + o (1))$ for
any given $\varepsilon$ and $\nu$.

Choosing $\varepsilon / 2$ in the place of $\varepsilon$ (and an arbitrary
$\nu$) shows that even a complexity of $L (1/3 + \varepsilon, o (1))$ is
achievable.
\medskip

\noindent {\bf Remark.} 
In the analysis, we have remained silent about the exact nature of the $o(1)$
terms. As long as a fixed number of them is involved, this does not pose any
problem. But at first sight, since Heuristic Result~\ref{prop:descent} is used
a non-constant number of times, one apparently needs to make the $o(1)$ terms
explicit to check that they do not sum up to something that is not
tending to zero. However, although the number of nodes in the tree of
Algorithm~\ref{alg:dlog} is in $g^{1/(3\varepsilon)}$, the $o(1)$ term is
the same for any given level in the tree, so that actually only the depth
of the tree is important for these $o(1)$-terms considerations. The depth
of the tree is in $1/(3\varepsilon)$, which is a constant, so that we
actually consider a constant number of $o(1)$ terms and need not make them
explicit.

\section{Extensions to wider families of curves}
\label{sec:extensions}

\subsection{Highly singular curves}

Consider the case where the curve has an equation of the appropriate
form, but with a genus that is much smaller than $nd$. Then letting
$g'=nd$, one may apply the exact same algorithms yielding an
$L(1/3+\varepsilon)$ complexity. However, the subexponential function is
now taken with respect to $q^{g'}$. This may still result in a subexponential
complexity in $q^g$, depending on the relation between $q$, $g$ and~$g'$.

\subsection{Different balancing between $n$ and $d$}

Here we consider the case where $n\approx g^\alpha$ and $d\approx
g^{1-\alpha}$ for $\alpha\in\left[\frac13, \frac12\right]$. We shall just give
an informal description of an algorithm that yields an $L(1/3)$ complexity
for the group structure.
Note that to obtain the claimed complexity without
$\varepsilon$, the bounds on $n$ and $d$ should resemble the ones we have in
Section~\ref{sec:grp}. For instance, bounds of the form
$n\le n_0g^\alpha\M^{-\alpha}$ and $d\le d_0g^{1-\alpha}\M^\alpha$ would
suffice. For the sake of better readability, we content ourselves with
approximate bounds.

Let us restrict to $C_{ab}$ curves for simplicity, and let us call
$P_\infty$ the unique place at infinity. We proceed as in
Algorithm~\ref{algo:grp}, but the functions we consider are of the more
general form:
$$ \varphi = a_0(X) + a_1(X)Y + \cdots + a_k(X)Y^k,$$
where the $a_i(X)$ have a degree bounded by $g^\beta$ and $k$ is taken of
the form $g^\gamma$, for some $\beta$ and $\gamma$ to be determined. Then
the divisor of $\varphi$ is of the form $E - (\deg E)P_\infty$, with $E$
effective of degree bounded by
$g^{\gamma+1-\alpha} + g^{\beta+\alpha}$.

Fix a smoothness bound of $g^{\beta + \gamma}$;
with the usual heuristic, one can find
$E$ that is smooth in time about $g^{\max(\alpha-\gamma,
(1-\alpha)-\beta)}$.
The consistency check that the sieving space must be
larger than the factor base yields the condition
$$ \beta+\gamma \ge \max(\alpha-\gamma, (1-\alpha)-\beta),$$
which gives $\beta+2\gamma\ge \alpha$ and $\gamma+2\beta\ge 1-\alpha$.
This in turn imposes that $\beta+\gamma \ge 1/3$. Therefore, in this
setting we can not hope to get something better than an $L(1/3)$
complexity. We now show that this complexity is achievable: taking
$\beta=2/3-\alpha$ and $\gamma = \alpha-1/3$, all the conditions are
verified, and the complexity is as announced.

In the particular case of $\alpha=1/3$, we recover $\beta=1/3$ and
$\gamma=0$, which corresponds to Algorithm~\ref{algo:grp}. In the other
extremal case $\alpha=1/2$, we get $\beta=\gamma=1/6$.

If $\alpha$ gets smaller than $1/3$, then the $L(1/3)$ complexity is not
achievable with this algorithm. In fact, for each value of $\alpha \in
[0,1/3]$, there is an $L(x)$ complexity with $x \in [1/3, 1/2]$, and
finally, for hyperelliptic curves one essentially recovers
Adleman-Demarrais-Huang's $L(1/2)$ algorithm.
\bigskip

All of this concerns only the group structure. For the special-Q descent
however, things get more complicated and the $L(1/3+\varepsilon)$
complexity is lost when $\alpha$ is bigger than $1/3$. More precisely,
the same kind of computations as above yields a complexity of
$L(\alpha+\varepsilon)$ for $\alpha \in [1/3,1/2]$.

\bibliographystyle{plain}
\bibliography{l13}

\begin{thebibliography}{10}

\bibitem{AdDeHu94}
L.~M. Adleman, J.~DeMarrais, and M.-D. Huang.
\newblock A subexponential algorithm for discrete logarithms over the rational
  subgroup of the jacobians of large genus hyperelliptic curves over finite
  fields.
\newblock In L.~Adleman and M.-D. Huang, editors, {\em ANTS-I}, volume 877 of
  {\em Lecture Notes in Comput. Sci.}, pages 28--40. Springer--Verlag, 1994.

\bibitem{BePo98}
R.~L. Bender and C.~Pomerance.
\newblock Rigorous discrete logarithm computations in finite fields via smooth
  polynomials.
\newblock In D.~A. Buell and J.~T. Teitelbaum, editors, {\em Computational
  Perspectives on Number Theory: Proceedings of a Conference in Honor of A.O.L.
  Atkin}, volume~7 of {\em Studies in Advanced Mathematics}, pages 221--232.
  American Mathematical Society, 1998.

\bibitem{Couveignes01}
J.-M. Couveignes.
\newblock Algebraic groups and discrete logarithm.
\newblock In {\em Public-key cryptography and computational number theory},
  pages 17--27. de Gruyter, 2001.

\bibitem{Diem06}
C.~Diem.
\newblock An index calculus algorithm for plane curves of small degree.
\newblock In F.~He{\ss}, S.~Pauli, and M.~Pohst, editors, {\em ANTS-VII},
  volume 4076 of {\em Lecture Notes in Comput. Sci.}, pages 543--557.
  Springer--Verlag, 2006.

\bibitem{Enge02}
A.~Enge.
\newblock Computing discrete logarithms in high-genus hyperelliptic {J}acobians
  in provably subexponential time.
\newblock {\em Math. Comp.}, 71:729--742, 2002.

\bibitem{EnGa02}
A.~Enge and P.~Gaudry.
\newblock A general framework for subexponential discrete logarithm algorithms.
\newblock {\em Acta Arith.}, 102:83--103, 2002.

\bibitem{EnSt02}
A.~Enge and A.~Stein.
\newblock Smooth ideals in hyperelliptic function fields.
\newblock {\em Math. Comp.}, 71:1219--1230, 2002.

\bibitem{Gaudry00}
P.~Gaudry.
\newblock An algorithm for solving the discrete log problem on hyperelliptic
  curves.
\newblock In B.~Preneel, editor, {\em Advances in Cryptology -- EUROCRYPT
  2000}, volume 1807 of {\em Lecture Notes in Comput. Sci.}, pages 19--34.
  Springer--Verlag, 2000.

\bibitem{GaThThDi07}
P.~Gaudry, E.~Thom{\'e}, N.~Th{\'e}riault, and C.~Diem.
\newblock A double large prime variation for small genus hyperelliptic index
  calculus.
\newblock {\em Math. Comp.}, 76:475--492, 2007.

\bibitem{Hache96}
G.~Hach{\'e}.
\newblock {\em Construction effective de codes g{\'e}om{\'e}triques}.
\newblock PhD thesis, Universit{\'e} de Paris VI, 1996.

\bibitem{HaMc89}
J.~L. Haffner and K.~S. McCurley.
\newblock A rigorous subexponential algorithm for computation of class groups.
\newblock {\em J. Amer. Math. Soc.}, 2(4):837--850, 1989.

\bibitem{Hess02}
F.~He{\ss}.
\newblock Computing {R}iemann-{R}och spaces in algebraic function fields and
  related topics.
\newblock {\em J. Symbolic Comput.}, 33:425--445, 2002.

\bibitem{Hess04}
F.~He{\ss}.
\newblock Computing relations in divisor class groups of algebraic curves over
  finite fields.
\newblock Preprint, 2004.

\bibitem{Manstavicius92}
E.~Manstavi\v{c}ius.
\newblock Semigroup elements free of large prime factors.
\newblock In F.~Schweiger and E.~Manstavi\v{c}ius, editors, {\em New Trends in
  Probability and Statistic}, pages 135--153, 1992.

\bibitem{MuStTh99}
V.~M\"uller, A.~Stein, and C.~Thiel.
\newblock Computing discrete logarithms in real quadratic congruence function
  fields of large genus.
\newblock {\em Math. Comp.}, 68(226):807--822, 1999.

\bibitem{Storjohann00}
A.~Storjohann.
\newblock {\em Algorithms for Matrix Canonical Forms}.
\newblock PhD thesis, Eidgen\"ossische Technische Hochschule Z\"urich, 2000.

\end{thebibliography}

\end{document}